# MAGNETIC PROPERTIES of SHORT PERIOD InGaMnAs/InGaAs SUPERLATTICES

J. SADOWSKI[a, b, c], R. MATHIEU[d], P. SVEDLINDH[d], J. KANSKI[b], M. KARLSTEEN[b], K. SWIATEK[c], J. Z. DOMAGALA[c]

[a] *Niels Bohr Institute, Copenhagen University, DK-2100 Copenhagen, Denmark;*
[b] *Chalmers University of Technology, SE-412 96 Göteborg, Sweden;*
[c] *Institute of Physics, Polish Academy of Sciences, Al. Lotnikow 32/46, 02-668 Warszawa, Poland*
[d] *Uppsala University, Department of Materials Science, SE-751 21 Uppsala, Sweden*

We have observed a paramagnetic-to-ferromagnetic phase transition in short period InGaMnAs/InGaAs superlattices. The thicknesses of magnetic InGaMnAs layers in the structures studied was chosen to be 4 or 8 molecular layers (12 Å or 24 Å). The nonmagnetic InGaAs spacer layers are 12 Å thick. The In content of InGaMnAs and InGaAs layers was chosen in such a way that magnetic layers were: deep potential wells, high potential barriers or shallow potential wells. For superlattices with 8 monolayer thick InGaMnAs magnetic layers and 4 monolayer thick InGaAs nonmagnetic spacers the temperatures of paramagnetic-to-ferromagnetic phase transition do not depend on the band offsets between InGaMnAs and InGaAs adjusted by the In content.

## 1. Introduction

InGaMnAs belongs to the family of III-Mn-V ferromagnetic semiconductors, which has been extensively studied in recent years by many research groups. The main interest in magnetic semiconductors nowadays is associated with the newly studied phenomena involving spins of charge carriers, which can potentially be applied to create the new kind of electronic (spintronic) devices [1]. The two ternary compounds from the low In and high In composition boundary, namely InMnAs [2, 3] and GaMnAs [4 – 8] have been studied for a couple of years and paramagnetic-to-ferromagnetic phase transition was found in both of them, with the maximum critical temperatures ($T_c$) of 50 K and 110 K, respectively. Recently it has been reported [9] that InGaMnAs with In content close to 50% exhibits enhanced temperatures of paramagnetic to ferromagnetic phase transition, in comparison to GaMnAs with the same Mn content, for Mn compositions exceeding 6%. The maximum $T_c$ for InGaMnAs with 50%In and 9%Mn was found to be above 100 K [9].

Our work concentrates on magnetism in short period InGaMnAs/InGaAs superlattice structures (SLs). We have already studied in details the short period superlattice structures containing GaMnAs magnetic layers and GaAs nonmagnetic spacers [10] and found the presence of paramagnetic-to ferromagnetic phase transition in the structures with very thin GaMnAs layers (22 Å). Neutron diffraction measurements demonstrated that in the superlattices of this kind the adjacent GaMnAs layers are ferromagnetically coupled [11, 12]. In short period GaMnAs/GaAs SLs we found that the temperature of ferromagnetic phase transition dependends on the thickness of nonmagnetic spacer layers. Superlattices containing InGaAs in the spacer layers and InGaMnAs in magnetic layers open two additional degrees of freedom concerning adjusting both the valence band offsets between magnetic layers and nonmagnetic spacers and the strain state of the system.

## 2. Sample preparation

As is well known III-Mn-V magnetic semiconductors can only be grown by a low temperature Molecular Beam Epitaxy (LT MBE) technique. Low temperature means the temperature in the range of 180 °C to 300 °C. The exact MBE growth temperature is dependent on Mn content since higher concentration of Mn ions entering the Ga sites demands lower growth temperatures [6, 8]. The other growth parameters such as As to Ga flux ratio are also vary important, since besides the growth temperature this flux ratio determines the concentrations of As antisite defects ($As_{Ga}$) in the III-Mn-V material [13]. As has been shown by several groups both theoretically [14, 15] and experimentally [16, 17] the concentration of $As_{Ga}$ is crucial for magnetic properties of GaMnAs, such as $T_c$ and saturation magnetization.

The SLs were grown on epiready GaAs(100) substrates. Before the LT GaMnAs growth the standard procedure of growing 0.1 – 0.5 micrometer thick, high temperature (HT) GaAs buffer layer has been followed. In the case of structures containing In either in the magnetic layer or in the nonmagnetic spacer a thick (0.3 – 1.5 micrometer), InGaAs buffer was grown at 500 °C – 530 °C. The composition of this buffer was chosen to match the lattice parameter of subsequently grown InGaMnAs/InGaAs SL. The growth of thick HT InGaAs buffer was then followed by a thin (0.01 – 0.2 micrometer) LT InGaAs buffer with the same In content. In the case of GaMnAs/GaAs superalttice only a thin LT GaAs buffer was deposited. As is well known the lattice constant of GaMnAs ($a_{GaMnAs}$) is slightly bigger then that of GaAs. The dependence of $a_{GaMnAs}$ on the Mn content depends on the growth conditions, namely growth temperature [18] and As to Ga flux ratio [19]. In the the MBE growth conditions used by us: the substrate temperature ($T_s$) was close to 230 °C and As to Ga flux ratio close to 2. We use arsenic dimmers (generated by valved cracker DCA effusion cell) for the MBE growth, which offers possibilities of precise control of As flux, difficult to achieve using the standard $As_4$ source. In our case extrapolation of the $a_{GaMnAs}$ dependence on Mn content gives the lattice constant of hypothetical zinc-blende MnAs compound of 5.90 Å. Assuming a linear dependence of GaMnAs lattice parameter on Mn content, which as we have found [19] is true only for Mn content higher then the content of $As_{Ga}$ defects, we get the lattice constant for GaMnAs containing 6% Mn equal to 5.67 Å. That means that the strain between $Ga_{0.94}Mn_{0.06}As$ and GaAs(100) substrate is equal to 0.3%. With this small strain there is possible to grow single GaMnAs films with the same in plane lattice constant as GaAs substrate up to the thickness of 2 micrometers. In connection to this it's clear that it is not necessary to deposit the relaxed buffer layer lattice matched to the GaMnAs/GaAs SL. In the case of InGaMnAs/InGaAs SLs that is not the case, since the In composition chosen for all the structures is equal to 50%. In all the samples the Mn content in magnetic layers was chosen to be 6%.

The growth of each SL structure was monitored by reflection high energy electron diffraction (RHEED) system. Growth rate measurements based on RHEED intensity oscillations were performed for: GaAs, InGaAs and GaMnAs prior each SL growth. Thus both the composition and thicknesses of the constituent SL layers have been carefully measured. At each interface a 30 s growth brake was used in order to smooth the surfaces of each of the SL constituent layers. The parameters of samples are listed in Table 1.

## 3. X-ray diffraction results

Since in the case of structures containing 50%In in magnetic layers or/and in nonmagnetic spacers the lattice mismatch between InGaMnAs and InGaAs to GaAs is high ( 7%) we have grown the thick, relaxed InGaAs buffer layers with the composition adjusted to the calculated lattice constant of the free standing (unstrained) SL structure. Fig. 1 shows the X-ray diffraction results for sample 1 – InGaMnAs / GaAs SL and sample 4 - GaMnAs / InGaAs SL. In the case of sample 1 (Fig. 1a) - the SL structure with 8 ML thick $(In_{0.5}Ga_{0.5})_{0.94}Mn_{0.06}As$ and 4 ML thick GaAs. The InGaAs buffer lattice matched to the SL structure is $In_{0.33}Ga_{0.67}As$. For sample 4 containing 8 ML thick $Ga_{0.94}Mn_{0.06}As$ and 4 ML thick $In_{0.5}Ga_{0.5}As$ the lattice matched buffer is $In_{0.167}Ga_{0.833}As$. The first peak at the left side to the GaAs substrate 004 Bragg reflection in both cases comes from HT InGaAs buffer. Performing reciprocal space map measurements for asymmetric (224) Bragg reflection we verified that for both samples this buffer is partially relaxed. Analyzing the RHEED diffraction patterns recorded during InGaAs buffer deposition we verified that after a few monolayers of coherently strained InGaAs wetting layer (8 monolayers in the case of $In_{0.167}Ga_{0.833}As$ and 4 monolayers for $In_{0.5}Ga_{0.5}As$) the growth mode changes from two to three dimensional (i.e. InGaAs quantum dots are formed), but shortly afterwards these dots are overgrown by InGaAs and two dimensional growth mode is recovered after less than 10 monolayers of InGaAs deposited. Reciprocal space maps showed that InGaAs buffer is partially relaxed and partially strained to the GaAs substrate. It can be seen in Fig. 1 that the perpendicular lattice parameter of InGaAs buffer is slightly larger than mean perpendicular lattice parameter of superlattice, since the angular position of InGaAs(004) Bragg reflection is slightly smaller than the angular position of zero order SL peak. The lattice constant of InGaAs for each SL was calculated to perfectly match the InGaAs buffer lattice constant on the assumption that InGaAs is fully relaxed, which is not the case. Nevertheless the lattice mismatch between InGaAs buffer and superlattice is very small in each case, so the superlattices are fully strained to the buffers. This has been confirmed by reciprocal space maps performed for asymmetric (224) reflections.

From all the samples listed in Tab.1, in most of the cases magnetic superlattice layers are under compressive strain. The only sample for which magnetic layers are in tensile strain state is sample 4 - the SL structure with 8 monolayers GaMnAs and 4 monolayers $In_{0.5}Ga_{0.5}As$ grown on $In_{0.167}Ga_{0.833}As$ buffer.

.

## 4. Magnetic properties

The temperature dependence of the field cooled magnetization of all samples was recorded in a Superconducting QUantum Interference Device (SQUID), using a small magnetic field. The magnetization (M) versus temperature (T) curves shown in Fig. 2 show that the ferromagnetic ordering temperatures are dependent mainly on the composition of magnetic layers, differing substantially for SL structures with GaMnAs and InGaMnAs layers. The main frame in Fig. 2 shows M vs. T curves for two superlattices with GaMnAs magnetic layers differing in the composition of spacer layers: GaAs or $In_{0.5}Ga_{0.5}As$ in samples 3 and 4, respectively. It can be seen that the composition of the 4 ML spacer layer has no significant influence on $T_c$. For both structures with GaMnAs magnetic layers $T_c$ values are close to 40 K. It should be noted however, that samples 3 and 4 differ also in the strain state of magnetic GaMnAs layers – compressive strain for sample 3 and tensile strain for sample 4.

The inset to Fig. 2 shows M vs. T curves for two superlattices with InGaMnAs magnetic layers, and GaAs (sample 1) or $In_{0.5}Ga_{0.5}As$ (sample 2) spacers. A magnetization versus temperature curve for a reference 700 Å thick single InGaMnAs layer (sample 5) is also shown. This is interesting to notice that Tc in superlattices with InGaMnAs is about 10 K lower than $T_c$ of single InGaMnAs layer with the same composition. The comparison of magnetization temperature dependence for these two superlattices may indicate the influence of the confinement of holes on magnetic properties. In the case of sample 1 magnetic InGaMnAs layers are potential wells for holes and nonmagnetic GaAs spacers are barriers. In the case of sample 2 the band offsets between InGaAs and InGaMnAs are much smaller than for sample 1 since they results only from the very small composition difference. As can be seen in Fig. 2 the M(T) curves for these two samples differ significantly. It should be noted however that also the strain state of magnetic InGaMnAs layers is different in both structures: InGaMnAs is under quite important compressive strain in the case of sample 1, whereas in the case of sample 2 the strain results only from the differences between lattice parameters of low temperature InGaMnAs/InGaAs layers and the high temperature $In_{0.5}Ga_{0.5}As$ buffer. More studies are needed to clarify the influence of both strain and valence band offsets between magnetic and nonmagnetic layers in SLs structures of this kind. We should stress however that to our knowledge our results are the first observation of paramagnetic to ferromagnetic phase transition in InGaMnAs based structures with thickness of magnetic layers as low as 4 and 8 molecular layers.

## 5. Conclusions

We have found a paramagnetic-to-ferromagnetic phase transition in the short period InGaMnAs/InGaAs superlattices. For the thickness of magnetic layers equal to 4 or 8 molecular layers (12 Å and 24 Å, respectively) and the thickness of nonmagnetic InGaAs spacer of 4 ML we observed a ferromagnetic phase transition in both type of structures: with magnetic layers being potential barriers or wells for valence band holes. The difference in the temperature dependence of magnetization in different type of structures can be associated both with difference in band offsets or different strain state of magnetic layers in the superlattice structures.

**Acknowledgements**

The present work is supported by grants from the Swedish Research Council and by the Nanometer Consortium at Lund University. The MBE work was carried out at BL41 at the Swedish National Synchrotron Radiation Laboratory MAX-lab.

| Sample number | Magnetic layer | Spacer layer | buffer | Number of repetitions |
|---|---|---|---|---|
| 1 | InGaMnAs(8 ML) | GaAs(4 ML) | $In_{0.33}Ga_{0.67}As$ | 200 |
| 2 | InGaMnAs(4 ML) | InGaAs(4 ML) | $In_{0.5}Ga_{0.5}As$ | 200 |
| 3 | GaMnAs(8 ML) | GaAs(4 ML) | - | 200 |
| 4 | GaMnAs(8 ML) | InGaAs(4 ML) | $In_{0.167}Ga_{0.833}As$ | 200 |
| 5 | InGaMnAs(240 ML) | - | $In_{0.5}Ga_{0.5}As$ | - |

Table 1. Parameters of InGaMnAs/InGaAs superlattices structures. In all the structures Mn content is equal to 6 at. %

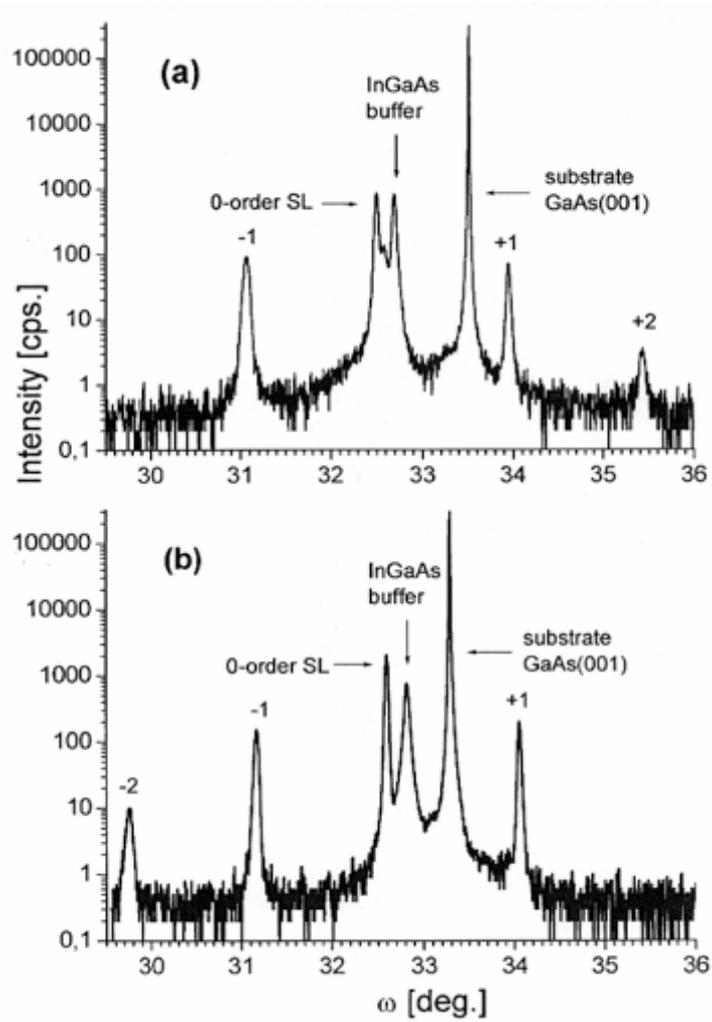

**Fig. 1.** X-ray rocking curves for (004) Bragg reflection for:

   (a) sample 1 - InGaMnAs(8 ML) / GaAs(4 ML) x 200 superlattice

   (b) sample 4 - GaMnAs(8 ML) / InGaAs(4 ML) x 200 superlattice

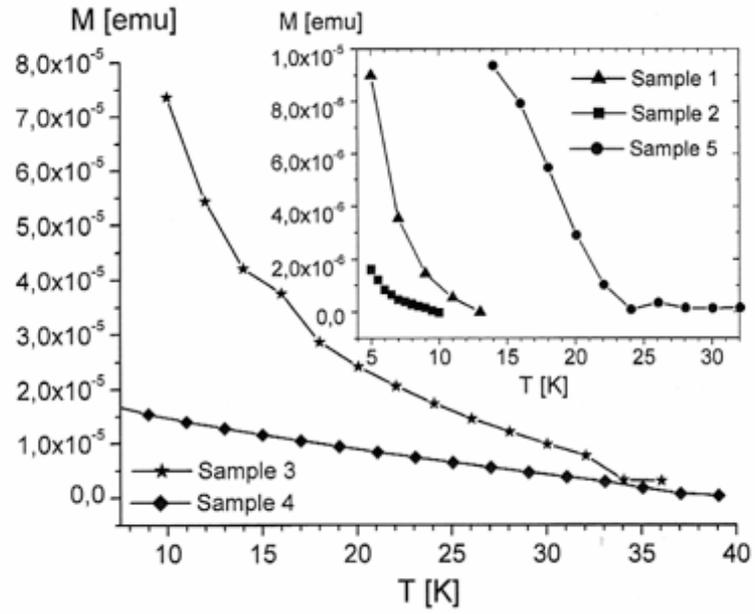

**Fig. 2.** Temperature dependence of the field cooled magnetization for SL structures with magnetic GaMnAs layers (main figure) and GaAs spacer (sample 3) or InGaAs spacer (sample 4); $H$=50 Oe. The inset shows magnetization for SLs with magnetic InGaMnAs layers (sample 1 and 2) and a reference single InGaMnAs (sample 5).